\documentclass[12pt]{iopart}

\usepackage{graphicx}
\usepackage{dsfont}
\usepackage{iopams}

\newcommand{\fig}[1]{Figure~\ref{#1}}

\newcommand{\be}{\begin{equation}}
\newcommand{\ee}{\end{equation}}

\begin{document}

\title{Recurrence of biased quantum walks on a line}

\author{M. \v Stefa\v n\'ak$^{(1)}$, T. Kiss$^{(2)}$ and I. Jex$^{(1)}$}
\address{$^{(1)}$ Department of Physics, FJFI \v CVUT v Praze, B\v
rehov\'a 7, 115 19 Praha 1 - Star\'e M\v{e}sto, Czech Republic}
\address{$^{(2)}$ Department of Nonlinear and Quantum Optics, Research
Institute for Solid State Physics and Optics, Hungarian Academy of
Sciences, Konkoly-Thege u.29-33, H-1121 Budapest, Hungary}


\date{\today}

\begin{abstract}
The P\'olya number of a classical random walk on a regular lattice is known to
depend solely on the dimension of the lattice. For one and two dimensions it equals one, meaning unit probability to return to the origin. This result is
extremely sensitive to the directional symmetry, any deviation
from the equal probability to travel in each direction results in a
change of the character of the walk from recurrent to transient.
Applying our definition of the P\'olya number to quantum walks on a
line we show that the recurrence character of quantum walks is more stable against bias. We determine the range of parameters for which biased quantum walks remain recurrent. We find that there exist genuine biased
quantum walks which are recurrent.
\end{abstract}

\maketitle

%%%%%%%%%%%%%%%%%%%%%%%%%%%%%%%%%%%%%%%%%%%%%%%%%%%%%%%%%%%%%%%

\section{Introduction}
\label{sec1}

Random walks are a popular topic in physics \cite{overview,hughes}. The popularity
stems from several sources. First, random walks are rather simple in their
formulation yet powerful in their application and allow to pinpoint the essential
physics involved in the studied processes. Next, the random walks are
one of the tools which allow to connect the microdynamics with the
macrobehaviour of large systems. Finally, random walks are quite
flexible and popular also outside physics to describe various phenomena. Hence it comes not as a surprise
that the first random walks have not been formulated within physics but to describe the alternation of share prices on the stock
exchange or the spreading of insects in a forest \cite{bachelier,chandrasekhar:1943}.

The study of random walks obtained a new stimulus when they were combined with quantum mechanics \cite{aharonov,konno:book}. Here the walker is thought to be a non-classical object enriched with wave attributes. The novel features of
quantum walks have been shown to be not only of theoretical interest
but to also have practical implications, especially for quantum algorithms \cite{kempe,shenvi:2003,aurel:2007,santha}. An important concept is the hitting time \cite{kempe:2005,krovi:2006a,krovi:2006b,magniez} which helps to point out the fundamental difference between classical and quantum walks allowing for algorithmic speed-up. One of the simplest non-trivial examples for a quantum walk is the one on a line \cite{nayak} which is closely related to the so called optical Galton board \cite{optical:galton}. Various aspects of one dimensional quantum walks have been analyzed \cite{tregenna,wojcik,knight,carteret,chandrashekar:2007,konno:2002}. Additional interesting effects, e.g. localisation, arise when one considers multi-state quantum walks \cite{1dloc1,1dloc,miyazaki,sato}.

One of the characteristics of the random walk on an infinite lattice is
expressed by the probability of the walker to return to its starting position, called the P\'olya number \cite{polya}. If the P\'olya number equals one
the walk is called recurrent, otherwise there is a non-zero probability that the walker never returns to its starting position. Such walks are called transient. The recurrence nature of the random walk is determined by the asymptotic behaviour of the probability at the origin \cite{revesz}. One finds that a random walk is transient if the probability at the origin decays faster than $t^{-1}$. The recurrent behaviour has been studied in great detail for classical random walks in dependence on the dimension and the topology of the lattice \cite{domb:1954,montroll:1964}.

Recently, we have extended the concept of P\'olya number to quantum walks \cite{prl}. In our definition we proposed a particular measurement scheme to minimize disturbance: each measurement in a series is carried out on a different member of an ensemble of equally prepared quantum systems. We have shown that the recurrence nature of the quantum walk, according to the above definition, is determined by the asymptotic behaviour of the probability at the origin in a similar way as in classical random walks. However, due to interference the asymptotics of the probability at the origin does not depend solely on the dimension of the lattice, but also on the coin operator and the initial coin state. Hence, one can find strikingly different recurrence behaviour for quantum walks compared to their classical counterpart \cite{pra}. Note that recurrence is meant here as the return to the origin which can be considered as a fractional recurrence from the point of view of the whole quantum state \cite{peres,chandra:recurrence} So far we have considered balanced walks, i.e. ones where there is no preference in direction for the walker and the step lengths are equal. For a large class of quantum walks this assumption does not hold and we wish to study the implications of unbalanced coins and unequal step lengths for the recurrence properties.

In the present paper we study biased quantum walks on the line
and compare their properties with their classical counterparts. As we briefly review in \ref{app:a}, recurrence of classical random walks is a consequence of the walk's symmetry. They are recurrent if and only if the mean value
of the position of the particle vanishes. This is due to the fact
that the spreading of the probability distribution of the
position is diffusive while the mean
value of the position propagates with a constant velocity. In contrast, for
quantum walks both the spreading of the probability distribution and
the propagation of the mean value are ballistic. We show that this allows for maintaining recurrence even when the symmetry is broken.

Our paper is organized as follows: In Section~\ref{sec2} we describe
the biased quantum walk on a line. In Section~\ref{sec3} we solve
the time evolution equations with the help of the Fourier
transformation. We find that the probability amplitudes can be
expressed in terms of integrals where time enters only in the
rapidly oscillating phase factor. This fact allows a straightforward
asymptotic analysis of the probability amplitudes by means of the
method of stationary phase. We perform this analysis in
Section~\ref{sec4}. Since the recurrence of the quantum walk is
determined by the asymptotics of the probability at the origin we
find a condition under which the biased quantum walk on a line is
recurrent. In Section~\ref{sec5} we analyze the recurrence of biased
quantum walks from a different perspective. We find that the
recurrence is related to the velocities of the peaks of the
probability distribution generated by the quantum walk. The explicit
form of the velocities leads us to the same condition derived in
Section~\ref{sec4}. Finally, in Section~\ref{sec6} we analyze the
formula for the mean value of the position of the particle derived
in \ref{app:b} in dependence of the parameters of the walk
and the initial state. We find that there exist genuine biased
quantum walks which are recurrent. Conclusions and outlook are
left for Section~\ref{sec7}.

%%%%%%%%%%%%%%%%%%%%%%%%%%%%%%%%%%%%%%%%%%%%%%%%%%%%%%%%%%%%%%%%%%%%%%%%%%%%%%%

\section{Description of the walk}
\label{sec2}

\begin{figure}
\begin{center}
\includegraphics[width=0.6\textwidth]{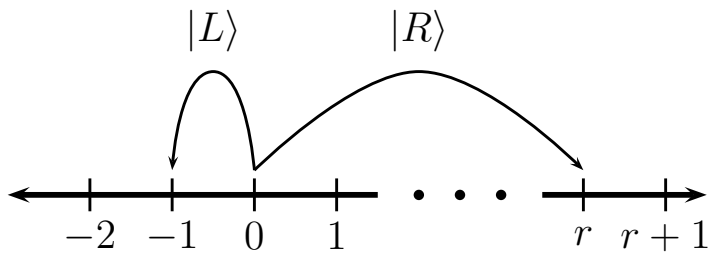}
\caption{Schematics of the biased quantum walk on a line. If the
coin is in the state $|R\rangle$ the particle moves to the right
to a point at distance $r$. With the coin state $|L\rangle$
the particle makes a unit length step to the left. Before the step
itself the coin state is rotated according to the coin operator
$C(\rho)$.}
\label{fig1}
\end{center}
\end{figure}

Let us consider biased quantum walks on a line where the particle has two possibilities --- jump to the right or to the left. Without loss of generality we restrict ourselves to biased quantum walks where the jump to the right is of the length $r$ and the jump to
the left has a unit size. We depict the biased quantum walk schematically in \fig{fig1}.

The Hilbert space of the particle has the form of the tensor product
\begin{equation}
{\cal H} = {\cal H}_P\otimes{\cal H}_C
\end{equation}
of the position space
\begin{equation}
{\cal H}_P = \ell^2(\mathds{Z}) = \textrm{Span}\left\{|m\rangle|\ m\in\mathds{Z}\right\},
\end{equation}
and the two dimensional coin space
\begin{equation}
{\cal H}_C = \mathds{C}^2 = \textrm{Span}\left\{|R\rangle,|L\rangle\right\}.
\end{equation}
A single step of the quantum walk is given by the propagator
\begin{equation}
U = S \left(I_P\otimes C\right).
\label{qw:time}
\end{equation}
Here $I_P$ denotes the unit operator acting on the position space $\mathcal{H}_P$. The displacement operator $S$ has the form
\begin{equation}
S  = \sum\limits_{m=-\infty}^{+\infty}|m+r\rangle\langle m|\otimes|R\rangle\langle R|+\sum\limits_{m=-\infty}^{+\infty}|m-1\rangle\langle m|\otimes|L\rangle\langle L|.
\end{equation}
The coin flip $C$ is in general an arbitrary unitary operator acting on the coin space $\mathcal{H}_C$ and is applied on the
coin state before the displacement $S$ itself. However, as has been discussed in \cite{tregenna} the probability distribution is not affected by the complex phases of the coin operator. Hence, it is sufficient to consider the one-parameter
family of coins
\begin{equation}
C(\rho) = \left(
            \begin{array}{cc}
              \sqrt{\rho} & \sqrt{1-\rho} \\
              \sqrt{1-\rho} & -\sqrt{\rho} \\
            \end{array}
          \right).
\label{coins}
\end{equation}
From now on we restrict ourselves to this family
of coins. The value of $\rho=1/2$ corresponds to the well known case of the Hadamard walk.

We write the initial state of the particle in the form
\begin{equation}
|\psi(0)\rangle \equiv
\sum\limits_{m=-\infty}^{+\infty}\sum_{i=R}^L\psi_i(m,0)|m\rangle\otimes|i\rangle.
\end{equation}
The state of the walker after $t$ steps is given by successive
application of the time evolution operator given by Eq.
(\ref{qw:time}) on the initial state
\begin{equation}
|\psi(t)\rangle \equiv
U^t|\psi(0)\rangle =
\sum\limits_{m=-\infty}^{+\infty}\sum_{i=R}^L\psi_i(m,t)|m\rangle\otimes|i\rangle.
\label{time:evol}
\end{equation}
The state of the particle is fully
determined by the set of two-component vectors
\begin{equation}
\psi(m,t)\equiv{\left(\psi_R(m,t),\psi_L(m,t)\right)}^T.
\label{prob:ampl}
\end{equation}
Here $\psi_{R(L)}(m,t)$ is the probability
amplitude to find the particle at position $m$ after $t$ steps with
the coin state $|R(L)\rangle$. The probability distribution
generated by the quantum walk is given by
\begin{eqnarray}
\nonumber P(m,t) & = & |\langle m,R|\psi(t)\rangle|^2+|\langle m,L|\psi(t)\rangle|^2 \\
\nonumber & = & |\psi_R(m,t)|^2+|\psi_L(m,t)|^2 = \|\psi(m,t)\|^2.\\
\end{eqnarray}

%%%%%%%%%%%%%%%%%%%%%%%%%%%%%%%%%%%%%%%%%%%%%%%%%%%%%%%%%%%%%%%%%%%%%%%%%%%%%%%

\section{Time evolution of the walk}
\label{sec3}

To obtain explicit and closed form expressions for the time
dependent state vector we rewrite the time evolution equation
(\ref{time:evol}) for the state vector $|\psi(t)\rangle$ into a set
of difference equations
\begin{eqnarray}
\nonumber \psi(m,t) & = & C_+(\rho)\psi(m-r,t-1)+\\
& & +C_-(\rho)\psi(m+1,t-1)
\label{time:evol2}
\end{eqnarray}
for the probability amplitude vectors $\psi(m,t)$. The form of the
matrices $C_\pm(\rho)$ follows from the matrix $C(\rho)$
\begin{equation}
C_+(\rho)  =  \left(
              \begin{array}{cc}
                \sqrt{\rho} & \sqrt{1-\rho} \\
                0 & 0 \\
              \end{array}
            \right),\qquad
C_-(\rho)  = \left(
              \begin{array}{cc}
                 0 & 0 \\
                \sqrt{1-\rho} & -\sqrt{\rho} \\
              \end{array}
            \right).
\end{equation}

The time evolution equations (\ref{time:evol2}) are greatly
simplified with the help of the Fourier transformation
\begin{equation}
\tilde{\psi}(k,t)\equiv\sum\limits_{m=-\infty}^{+\infty}\psi(m,t)e^{i mk},
\label{qw:ft}
\end{equation}
where the momentum $k$ is a continuous parameter ranging from $-\pi$ to $\pi$. The new function
$\tilde{\psi}(k,t)$ is square integrable on a unit circle.

The time evolution in the Fourier picture turns into a single difference equation
\begin{equation}
\tilde{\psi}(k,t) = \widetilde{U}(k)\tilde{\psi}(k,t-1),
\label{qw:te:fourier}
\end{equation}
where the propagator has the form
\begin{equation}
\widetilde{U}(k) \equiv \left(
                                            \begin{array}{cc}
                                              \sqrt{\rho}e^{ikr} & \sqrt{1-\rho}e^{ikr} \\
                                              \sqrt{1-\rho}e^{-ik} & -\sqrt{\rho}e^{-ik} \\
                                            \end{array}
                                          \right).
\label{teopF}
\end{equation}
The solution of (\ref{qw:te:fourier}) is
straightforward. We find
\begin{equation}
\tilde{\psi}(k,t) =\widetilde{U}^t(k)\tilde{\psi}(k,0),
\end{equation}
where $\tilde{\psi}(k,0)$ is the Fourier transformation of the initial
state. We restrict ourselves to the situation where the particle is
initially localized at the origin as dictated by the nature of
the problem we wish to study. As follows from (\ref{qw:ft}) the
Fourier transformation $\tilde{\psi}(k,0)$ of such an initial
condition is equal to the initial state of the coin
\begin{equation}
\tilde{\psi}(k,0) = \left(
                      \begin{array}{c}
                        \psi_R(0,0) \\
                        \psi_L(0,0) \\
                      \end{array}
                    \right),
\end{equation}
which we denote by $\psi$. Since $\psi$ can be an arbitrary
normalized complex two-component vector we parameterize it
by two parameters $a\in[0,1]$ and $\varphi\in[0,2\pi)$ in the
form
\begin{equation}
\psi = \left(
         \begin{array}{c}
           \sqrt{a} \\
           \sqrt{1-a}e^{i\varphi} \\
         \end{array}
       \right).
\label{psi:init}
\end{equation}
To evaluate the powers of the propagator $\widetilde{U}(k)$
it is convenient to diagonalize it. Since the propagator is
unitary its eigenvalues have the form $e^{i\omega_{1,2}}$ where the
phases read
\begin{eqnarray}
\nonumber \omega_1(k) & = & \frac{r-1}{2}k+\arcsin\left(\sqrt{\rho}\sin\left(\frac{r+1}{2}k\right)\right),\\
\omega_2(k) & = & \frac{r-1}{2}k -\pi-\arcsin\left(\sqrt{\rho}\sin\left(\frac{r+1}{2}k\right)\right)
\label{omega}
\end{eqnarray}
We denote the corresponding eigenvectors by $v_{1,2}(k)$. We
give their explicit form in the \ref{app:b}. With this notation we write
the solution of the time evolution equation in the Fourier picture
in the form
\begin{equation}
\widetilde{\psi}(k,t) = \sum_{j=1}^2
e^{i\omega_j(k)t}\left(v_j(k),\psi\right)v_j(k).
\end{equation}
Here $( , )$ means scalar product in the coin space. Finally, we obtain the solution in position representation by
performing the inverse Fourier transformation
\begin{eqnarray}
\nonumber \psi(m,t)  =  \int_{-\pi}^\pi\frac{dk}{2\pi}\ \widetilde{\psi}(k,t)\ e^{-imk} =  \sum_{j=1}^2\int_{-\pi}^\pi\frac{dk}{2\pi}e^{i(\omega_j(k)t-mk)}\ \left(v_j(k),\psi\right)v_j(k).\\
\label{inv:f}
\end{eqnarray}

%%%%%%%%%%%%%%%%%%%%%%%%%%%%%%%%%%%%%%%%%%%%%%%%%%%%%%%%%%%%%%%%%%%%%%%%%%%%%%%

\section{Asymptotics of the quantum walk and recurrence}
\label{sec4}

\begin{figure}[h]
\begin{center}
\includegraphics[width=0.5\textwidth]{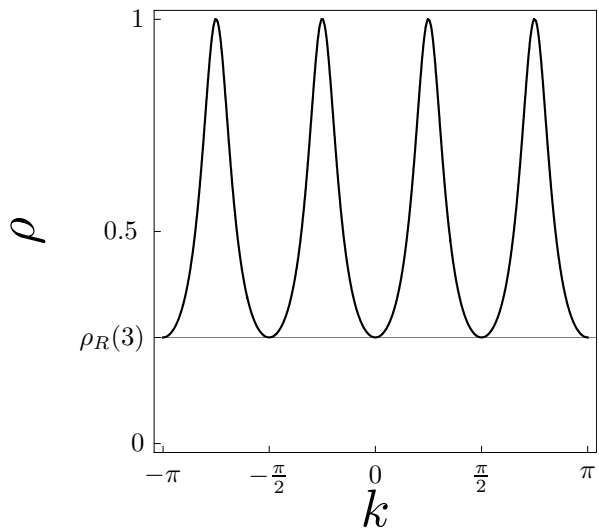}
\caption{The existence of stationary points of the phases
$\omega_{1,2}(k)$ in dependence on the parameter $\rho$ and
a fixed step length $r$. We plot the implicit functions $\omega_{1,2}'(k)\equiv
0$ for $r=3$. The plot indicates that for
$\rho<\rho_R(3)=\frac{1}{4}$ the phases $\omega_{1,2}(k)$ do
not have any saddle points. Consequently, the probability amplitude
at the origin decays fast and such biased quantum walk on a line is
transient. For $\rho\geq\rho_R(3)$ the saddle points exist and the
quantum walk is recurrent.}
\label{fig2}
\end{center}
\end{figure}

To determine the recurrence nature of the biased quantum walk we
have to analyze the asymptotic behaviour of the probability at the
origin \cite{prl}. Exploiting (\ref{inv:f}) the amplitude at the origin
reads
\begin{equation}
\psi(0,t) = \sum_{j=1}^2\int_{-\pi}^\pi\frac{dk}{2\pi}e^{i\omega_j(k)t}\
\left(v_j(k),\psi\right)v_j(k),
\label{psi:0}
\end{equation}
which allows us to find the asymptotics of the probability at the origin
with the help of the method of stationary phase \cite{statphase}.
The important contributions to the integrals in (\ref{psi:0}) arise
from the stationary points of the phases (\ref{omega}). We find that
the derivatives of the phases $\omega_{1,2}(k)$ are
\begin{eqnarray}
\nonumber \omega_1'(k) & = & \frac{r-1}{2}+\frac{\sqrt{\rho}(r+1)\cos\left(k\frac{r+1}{2}\right)}{\sqrt{4+2\rho\left[\cos(k(r+1))-1\right]}},\\
\nonumber \omega_2'(k) & = & \frac{r-1}{2}-\frac{\sqrt{\rho}(r+1)\cos\left(k\frac{r+1}{2}\right)}{\sqrt{4+2\rho\left[\cos(k(r+1))-1\right]}}.\\
\label{phase:der}
\end{eqnarray}
Using the method of stationary phase we find that the
amplitude will decay slowly - like $t^{-\frac{1}{2}}$, if at least
one of the phases has a vanishing derivative inside the integration
domain. Solving the equations $\omega_{1,2}'(k) = 0$ we find
that the possible saddle points are
\begin{equation}
k_0 = \pm\frac{2}{r+1}\arccos\left(\pm\sqrt{\frac{(1-\rho)(r-1)^2}{4\rho r}}\right).
\label{k0}
\end{equation}
The saddle points are real valued
provided the argument of the arcus-cosine in (\ref{k0}) is less or
equal to unity
\begin{equation}
\frac{(1-\rho)(r-1)^2}{4\rho r} \leq 1.
\end{equation}
This inequality leads us to the condition for the biased quantum
walk on a line to be recurrent
\begin{equation}
\rho_R(r) \geq \left(\frac{r-1}{r+1}\right)^2.
\label{crit:rec}
\end{equation}
We illustrate this result in \fig{fig2} for a particular choice of the walk parameter $r=3$.

Our simple result proves that there is an intimate nontrivial
link between the length of the step of the walk and the bias of the
coin. The parameter of the coin $\rho$ has to be at least equal to
a factor determined by the size of the step to the right $r$ for the
walk to be recurrent. We note that the recurrence nature of the
biased quantum walk on a line is determined only by the parameters
of the walk itself, i.e. the coin and the step, not by the initial
conditions. The parameters of the initial state $a$ and $\varphi$
have no effect on the rate of decay of the probability at the
origin.

%%%%%%%%%%%%%%%%%%%%%%%%%%%%%%%%%%%%%%%%%%%%%%%%%%%%%%%%%%%%%%%%%%%%%%%%%%%%%%%

\section{Recurrence of a quantum walk and the velocities of the peaks}
\label{sec5}

We can determine the recurrence nature of the biased quantum walk on
a line from a different point of view. This approach is based
on the following observation. The well known shape of the
probability distribution generated by the quantum walk consists of
two counter-propagating peaks. In between
the two dominant peaks the probability is roughly independent of $m$ and decays like $t^{-1}$.
On the other hand, outside the decay is exponential as we depart from the peaks. As it has been found in \cite{nayak} the position of the peaks varies linearly with the number of steps. Hence, the peaks propagate with constant velocities, say $v_L$ and $v_R$. For the
biased quantum walk to be recurrent the origin of the walk has to
remain in between the two peaks for all times. In other words, the
biased quantum walk on a line is recurrent if and only if the
velocity of the left peak is negative and the velocity of the right
peak is positive.

\begin{figure}[h]
\begin{center}
\includegraphics[width=0.7\textwidth]{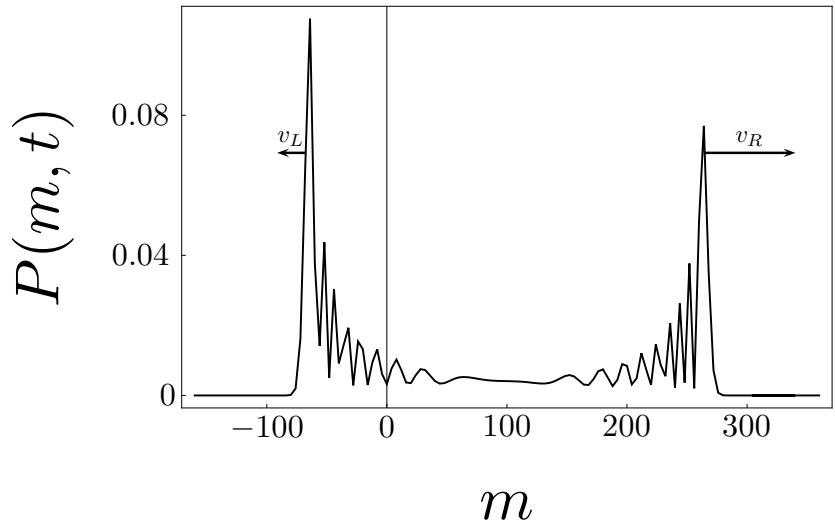}
\caption{Velocities of the left and right peak of the probability distribution generated by the biased quantum walk on a line and the recurrence.
We have chosen the parameters $r=3$, $a=\rho=\frac{1}{\sqrt{2}}$ and $\varphi = \pi$. The left peak propagates with the velocity $v_L\approx -0.68$,
the velocity of the right peak is $v_R\approx 2.68$. In between the two peaks the probability distribution behaves like $t^{-1}$ while outside the
decay is exponential. Since the velocity $v_L$ is negative the origin of the walk remains in between the left and right peak. Consequently, this quantum
walk is recurrent.}
\label{fig3}
\end{center}
\end{figure}

The velocities of the left and right peaks are easily determined. We
rewrite the formula (\ref{inv:f}) for the probability amplitude
$\psi(m,t)$ into the form
\begin{equation}
\psi(m,t) = \sum_{j=1}^2\int_{-\pi}^\pi\frac{dk}{2\pi}e^{i(\omega_j(k)-\alpha
k)t}\ \left(v_j(k),\psi\right)v_j(k),
\end{equation}
where we have introduced $\alpha = \frac{m}{t}$. Due to the fact that we now
concentrate on the amplitudes at the positions $m\sim t$ we have to
consider modified phases
\begin{equation}
\widetilde{\omega}_j(k) = \omega_j(k)-\alpha k.
\end{equation}
The peak occurs at such a position
$m_0$ where both the first and the second derivatives of
$\widetilde{\omega}_j(k)$ vanishes. The velocity of the peak is
thus $\alpha_0 = \frac{m_0}{t}$. Hence, solving the equations
\begin{eqnarray}
\nonumber \widetilde{\omega}_1'(k) & = & \frac{r-1}{2}+\frac{\sqrt{\rho}(r+1)\cos\left(k\frac{r+1}{2}\right)}{\sqrt{4+2\rho\left[\cos(k(r+1))-1\right]}} - \alpha = 0 ,\\
\nonumber \widetilde{\omega}_2'(k) & = & \frac{r-1}{2}-\frac{\sqrt{\rho}(r+1)\cos\left(k\frac{r+1}{2}\right)}{\sqrt{4+2\rho\left[\cos(k(r+1))-1\right]}} - \alpha = 0,\\
\nonumber \widetilde{\omega}_1''(k) & = & -\widetilde{\omega}_2''(k) = \frac{(\rho-1)\sqrt{\rho}(r+1)^2\sin\left(k\frac{r+1}{2}\right)}{\sqrt{2}\left[2-\rho+\rho\cos(k(r+1))\right]^\frac{3}{2}} = 0,\\
\end{eqnarray}
for $\alpha$ determines the velocities of the left and right peak $v_{L,R}$. The third equation is independent of $\alpha$ and we easily find the
solution
\begin{equation}
k_0 = \frac{4n\pi}{r+1},\ k_0=\frac{2\pi(2n+1)}{r+1},\ n\in\mathds{Z}.
\end{equation}
Inserting this $k_0$ into the first two equations we find the velocities of the left and right peak
\begin{eqnarray}
\nonumber v_L & = & \frac{r-1}{2}-\frac{r+1}{2}\sqrt{\rho}\\
v_R & = & \frac{r-1}{2}+\frac{r+1}{2}\sqrt{\rho}.
\label{velocities}
\end{eqnarray}
We illustrate this result in \fig{fig3} where we show the probability distribution generated by the quantum walk for the particular choice of the parameters $r = 3,\ \rho = \frac{1}{\sqrt{2}}$. The initial state was chosen according to $a = \frac{1}{\sqrt{2}}$ and $\varphi = \pi$. Since the velocity of the left peak $v_L$ is negative this biased quantum walk is recurrent.

The peak velocities have two contributions. One is identical
and independent of $\rho$, the second is a product of $r$ and $\rho$
and differs in sign for the two velocities. The obtained results indicate that biasing the walk by having
the size of the step to the right equal to $r$ results in dragging
the whole probability distribution towards the direction of
the larger step. This is manifested by the term $\frac{r-1}{2}$
which appears in both velocities $v_{L,R}$ with the same sign. On
the other hand the parameter of the coin $\rho$ does not bias the
walk. As we can see from the second terms entering the velocities it
rather influences the rate at which the walk spreads.

As we have discussed above the biased quantum walk on a line is recurrent if and only if $v_L$ is negative and $v_R$ is positive. The form of the
velocities (\ref{velocities}) implies that this condition is satisfied if and only if the criterion (\ref{crit:rec}) is fulfilled.

%%%%%%%%%%%%%%%%%%%%%%%%%%%%%%%%%%%%%%%%%%%%%%%%%%%%%%%%%%%%%%%%%%%%%%%%%%%%%%%

\section{Mean value of the biased quantum walk and recurrence}
\label{sec6}

As we discuss in the \ref{app:a} the classical random
walks are recurrent if and only if the mean value of the position
vanishes. We now show that this is not true for biased quantum
walks, i.e. there exist biased quantum walks on a line which are
recurrent but cannot produce probability distribution with zero mean
value. This is another unique feature of quantum walks
compared to the classical ones.

In the \ref{app:b} we derive the following formula
for the position mean value
\begin{eqnarray}
\nonumber \left\langle\frac{x}{t}\right\rangle & \approx & (1-\sqrt{1-\rho})(a(r+1)-1)+\\
\nonumber & & +\frac{\sqrt{a(1-a)}(1-\sqrt{1-\rho})(1-\rho)(r+1)\cos\varphi}{\sqrt{\rho(1-\rho)}}\\
& & +\frac{r-1}{2}\sqrt{1-\rho}+O(t^{-1}).
\label{mean}
\end{eqnarray}
We see that for quantum walks the mean value is affected by both the
fundamental walk parameters through  $r$ and $\rho$ and the
initial state parameters $a$ and $\varphi$. The mean value is
typically non-vanishing even for unbiased quantum walks (
with $r=1$ ). However, one easily finds \cite{tregenna} that the initial
state with the parameters $a=1/2$ and $\varphi=\pi/2$ results in a
symmetric probability distribution with zero mean independent of the
coin parameter $\rho$. Indeed, the quantum walks with $r=1$, i.e.
with equal steps to the right and left, do not intrinsically
distinguish left from right. On the other hand the quantum walks
with $r>1$ treat the left and right direction in a different way.
Nevertheless, one can always find for a given $r$ a coin parameter
$\rho_0$ such that for all $\rho\geq\rho_0$ the quantum walk can
produce a probability distribution with zero mean value. This is
impossible for quantum walks with $\rho<\rho_0$ and we will call
such quantum walks genuine biased.

Let us now determine the minimal value of $\rho$ for a given $r$ for
which mean value vanishes. We first find the parameters of the
initial state $a$ and $\varphi$ which minimizes the mean value.
Clearly the term on the second line in (\ref{mean}) reaches the
minimal value for $\varphi_0=\pi$. Differentiating the resulting
expression with respect to $a$ and setting the derivative equal to
zero gives us the condition
\begin{equation}
2 + \frac{(2a-1)
\sqrt{\rho(1-\rho)}}{\rho\sqrt{a(1-a)}} = 0
\end{equation}
on the minimal mean value with respect to $a$. This relation is satisfied for
$a_0=\frac{1}{2}(1-\sqrt{\rho})$. The resulting formula for the mean
value reads
\begin{equation}
\left\langle\frac{x}{t}\right\rangle_{a_0,\varphi_0}
= \frac{r-1}{2}+\frac{\left(1-\sqrt{1-\rho}-\rho\right) (1+r)}{2
\sqrt{(1-\rho) \rho}}.
\label{mean:min}
\end{equation}
This expression vanishes for
\begin{equation}
\rho_0(r) = \left(\frac{r^2 - 1}{r^2 + 1}\right)^2.
\label{rho:0}
\end{equation}
Since (\ref{mean:min}) is a decreasing function of
$\rho$ the mean value is always positive for $\rho<\rho_0$
independent of the choice of the initial state. For $\rho>\rho_0$
one can achieve zero mean value for different combination of the
parameters $a$ and $\varphi$.

The formula (\ref{rho:0}) is reminiscent of the condition
(\ref{crit:rec}) for the biased quantum walk on a line to be
recurrent. However, $r$ is in (\ref{rho:0}) replaced by $r^2$.
Therefore we find the inequality $\rho_R<\rho_0$. Hence, the quantum
walks with the coin parameter $\rho_R<\rho<\rho_0$ are recurrent but
cannot produce a probability distribution with zero mean value. We
conclude that there are genuine biased quantum walks which are
recurrent in contrast to situations found for classical walks.

%%%%%%%%%%%%%%%%%%%%%%%%%%%%%%%%%%%%%%%%%%%%%%%%%%%%%%%%%%%%%%%%%%%%%%%%%%%%%%%

\section{Conclusions}
\label{sec7}

We have analyzed one dimensional biased quantum walks. Classically, the bias leading to a non-zero mean value of the particle's position can be introduced in two ways --- unequal step lengths or unfair coin. In contrast, for quantum walks on a line the initial state can introduce bias for any coin. On the other hand, for symmetric initial state modifying only the unitary coin operator while keeping the equal step lengths will not introduce bias. Finally, the bias due to unequal step lengths may be compensated for by the choice of the coin operator for some initial conditions. For this reason we have introduced the concept of the genuinely biased quantum walk for which there does not exists any initial state leading to vanishing mean value of the position.

\begin{figure}[h]
\begin{center}
\includegraphics[width=0.6\textwidth]{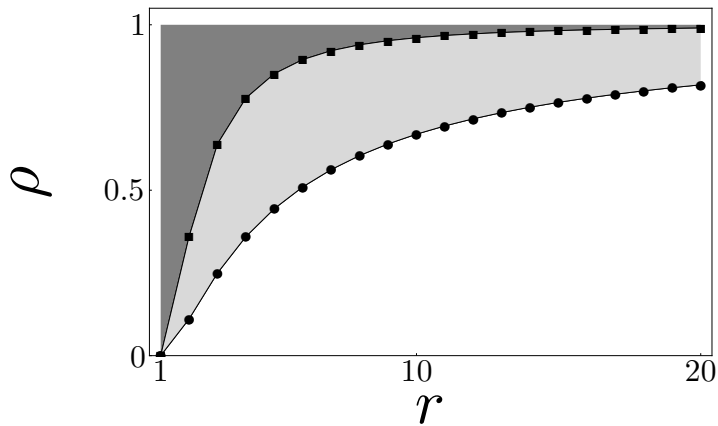}
\caption{"Phase diagram" of biased quantum walks on a line. The horizontal axis represents the length of the step to the right $r$ and the vertical axis shows the coin parameter $\rho$. The dotted line corresponds to the recurrence criterion (\ref{crit:rec}), while the squares represent the condition (\ref{rho:0}) on the zero mean value of the particle's position. The quantum walks in the white area are transient and genuine biased. In between the two curves (light gray area) we find quantum walks which are recurrent but still genuine biased. The quantum walks in the dark gray area are recurrent and for a particular choice of the initial state they can produce probability distribution with vanishing mean value.}
\label{fig4}
\end{center}
\end{figure}

We have determined the conditions under which one dimensional biased quantum walks are recurrent. This together with the condition of being genuinely biased give rise to three different regions in the parameter space which we depict as a "phase diagram" in \fig{fig4}.

The presented results allow for generalization to biased quantum walks in higher dimensions assuming we keep the coin operator
in a tensorial form. For non-factorizable coin operators in higher dimensions it remains an open question when they are recurrent or transient.

%%%%%%%%%%%%%%%%%%%%%%%%%%%%%%%%%%%%%%%%%%%%%%%%%%%%%%%%%%%%%%%%%%%%%%%%%%%%%%%%%

\ack

The financial support by MSM 6840770039, M\v SMT LC 06002, the Czech-Hungarian cooperation project (KONTAKT,CZ-10/2007) and by the Hungarian
Scientific Research Fund (T049234) is gratefully acknowledged.

%%%%%%%%%%%%%%%%%%%%%%%%%%%%%%%%%%%%%%%%%%%%%%%%%%%%%%%%%%%%%%%%%%%%%%%%%%%%%%%

\begin{appendix}

\section{Recurrence of classical biased random walk on a line}
\label{app:a}

\begin{figure}[h]
\begin{center}
\includegraphics[width=0.6\textwidth]{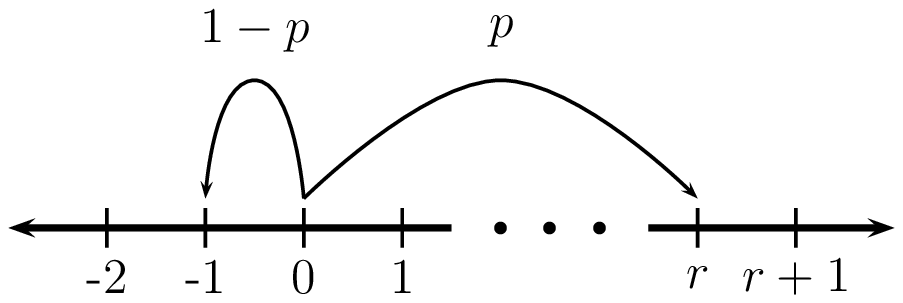}
\caption{Schematics of the biased random walk on a line. The particle can move to the right by a distance $r$ with the probability $p$.
The length of the step to the left is unity and the probability of this step is $1-p$.}
\label{fig5}
\end{center}
\end{figure}

Classical random walks on a line can be biased in two ways - the step in one direction is greater than in the other one and the probability of
the step to the right is different from the probability of the step to the left (see \fig{fig5}).

Consider a random walk on a line such that the particle can make a
jump of length $r$ to the right with probability $p$ or make a unit
size step to the left with probability $1-p$. The random walk is
recurrent if and only if the probability to find the particle at the
origin at any time instant $t$ does not decays faster than $t^{-1}$.
This probability is easily found to be expressed by the binomial
expression
\begin{equation}
P_0(t) = (1-p)^{\frac{t r}{r+1}}p^{\frac{t}{r+1}}{t\choose \frac{t r}{r+1}}.
\end{equation}
With the help of the Stirling's formula
\begin{equation}
n! \approx \sqrt{2\pi n}\left(\frac{n}{e}\right)^n
\end{equation}
we find the asymptotic behaviour of the probability at the origin
\begin{equation}
P_0(t)\approx \frac{r+1}{\sqrt{2\pi r t}}\left[(1-p)^{\frac{r}{r+1}}p^{\frac{1}{r+1}}\frac{r+1}{r^\frac{r}{r+1}}\right]^t.
\end{equation}
The asymptotics of the probability $P_0(t)$ therefore depends on
the value of
\begin{equation}
q = (1-p)^{\frac{r}{r+1}}p^{\frac{1}{r+1}}\frac{r+1}{r^\frac{r}{r+1}}.
\end{equation}
Since $q\leq 1$ the probability $P_0(t)$ decays exponentially
unless the inequality is saturated. Hence, the random walk is
recurrent if and only if $q$ equals unity. This condition is
satisfied for
\begin{equation}
\label{rw:cond}
p = \frac{1}{r+1},
\end{equation}
i.e. the probability of the step to the right has to be inversely
proportional to the length of the step.

\begin{figure}[h]
\begin{center}
\includegraphics[width=0.6\textwidth]{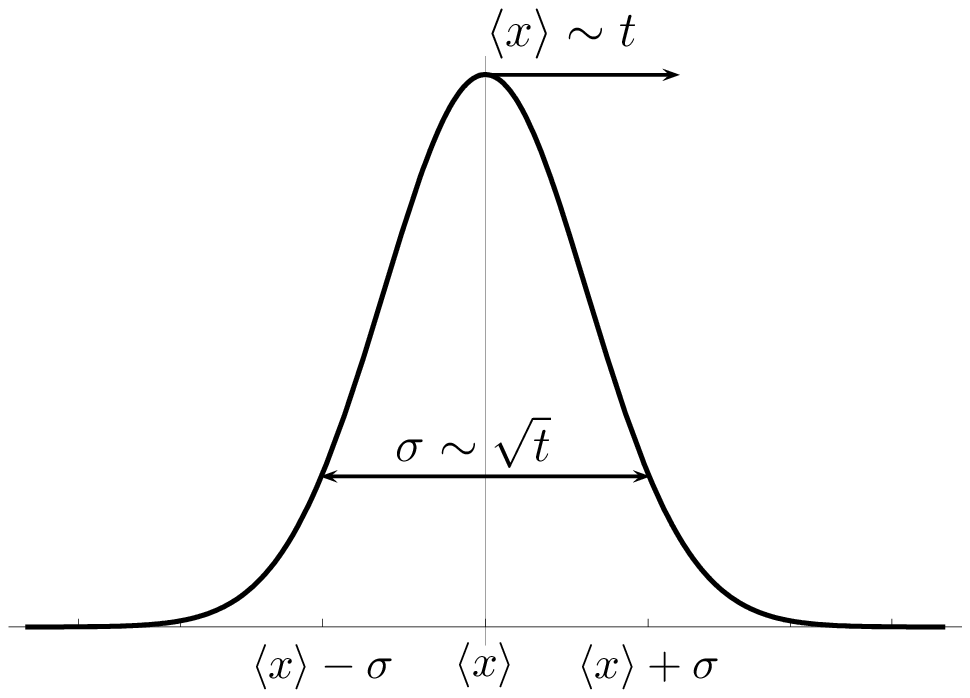}
\caption{Spreading of the probability distribution versus the motion
of the mean value of a biased classical random walk on a line. While the
spreading is diffusive ($\sigma\sim\sqrt{t}$) the mean value
propagates with a constant velocity ($\langle x\rangle\sim t$). The
probability inside the $\sigma$ neighborhood of the mean value
$\langle x\rangle$ behaves like $t^{-\frac{1}{2}}$. On the other
hand, as we go away from the $\sigma$ neighbourhood the decay is exponential.
Hence, if the mean value $\langle x\rangle$ does not vanish the
origin of the walk leaves the $\sigma$ neighborhood of the mean
value. In such a case the probability at the origin decays
exponentially and the walk is transient.}
\label{fig6}
\end{center}
\end{figure}

This result can be well understood from a different point of view, as we illustrate in \fig{fig6}.
The spreading of the probability distribution is diffusive, i.e.
$\sigma\sim\sqrt{t}$. The probability in the $\sigma$ neighborhood
of the mean value $\langle x\rangle$ behaves like $t^{-\frac{1}{2}}$
while outside this neighborhood the probability decays
exponentially. Therefore for the random walk to be recurrent the
origin must lie in this $\sigma$ neighborhood for all times $t$.
However, if the random walk is biased the mean value of the position
$\langle x\rangle$ varies linearly in time, thus it is a faster
process than the spreading of the probability distribution. In such
a case the origin would lie outside the $\sigma$ neighborhood of the mean
value after a finite number of steps leading to the exponential
asymptotic decay of the probability at the origin $P_0(t)$. Hence,
the random walk is recurrent if and only if the mean value of the
position equals zero. Since the individual steps are independent of
each other the mean value after $t$ steps is simply a $t$ multiple
of the mean value after single step, i.e.
\begin{equation}
\langle x (t)\rangle = t \langle x(1)\rangle = t\left[p(r+1)-1\right].
\end{equation}
We find that the mean value equals zero if and only if the condition (\ref{rw:cond})
holds.

%%%%%%%%%%%%%%%%%%%%%%%%%%%%%%%%%%%%%%%%%%%%%%%%%%%%%%%%%%%%%%%%%%%%%%%%%%%%%%%

\section{Mean value of the particle's position for a quantum walk on a line}
\label{app:b}

In this Appendix we find the explicit form of the position mean
value of the particle. With the help of the weak limit theorem
\cite{Grimmett} we express the mean value after $t$ steps in the
form
\begin{equation}
\left\langle \frac{x}{t}\right\rangle \approx \sum_{j=1}^2\int_{-\pi}^\pi\frac{dk}{2\pi}\ \omega_j'(k)\ \left(v_j(k),\psi\right)v_j(k),
\end{equation}
up to the corrections of the order $O(t^{-1})$. Here $v_j(k)$ are eigenvectors of the
unitary propagator $\widetilde{U}(k)$, $\omega_j'(k)$ are the
derivatives of the phases of the corresponding eigenvalues and
$\psi$ is the initial state expressed in (\ref{psi:init}). The
derivatives of the phases are given in (\ref{phase:der}). We express
the eigenvectors in the form
\begin{eqnarray}
\nonumber v_1(k) & = & n_1(\rho,k)\left(\sqrt{1-\rho}, -\sqrt{\rho} + e^{i(\omega_1(k)-rk)}\right)^T,\\
\nonumber v_2(k) & = & n_2(r,k)\left(\sqrt{1-\rho}, -\sqrt{\rho} + e^{i(\omega_2(k)-rk)}\right)^T.\\
\end{eqnarray}
The normalization factors of the eigenvectors read
\begin{eqnarray}
\nonumber n_1(u) & = & 2-2\sqrt{\rho}\cos\left(u-\arcsin\left[\sqrt{\rho}\sin u\right]\right),\\
\nonumber n_2(u) & = & 2+2\sqrt{\rho}\cos\left(u+\arcsin\left[\sqrt{\rho}\sin u\right]\right),\\
\end{eqnarray}
where we denote $u=\frac{k(r+1)}{2}$ to shorten the notation. The mean value is thus given by the following integral
\begin{equation}
\left\langle \frac{x}{t}\right\rangle \approx \int\limits_0^{(r+1)\pi}\frac{f(a,\varphi,\rho,r,u)du}{2(r+1)\pi \left[1 +\sqrt{\rho}\cos u_1\right] \left[1-\sqrt{\rho} \sin u_2\right]}+O(t^{-1}),
\end{equation}
where
\begin{equation}
u_1  =  u + \arcsin(\sqrt{\rho}\sin u),\quad u_2  =  u + \arccos(\sqrt{\rho}\sin u),
\end{equation}
and the numerator reads
\begin{eqnarray}
\nonumber f(a,\varphi,\rho,r,u) & = & (1-\rho) \left[r-1 + \rho \left(a + r (a-1)\right)\left(1+\cos(2u)\right)+\right.\\
\nonumber & & \left.+ \sqrt{a(1-a)}\sqrt{\rho(1-\rho)} (r+1) \left(\cos{\varphi}+\cos(\varphi+2u)\right)\right].\\
\end{eqnarray}
Performing the integrations we arrive at the result
\begin{eqnarray}
\nonumber \left\langle\frac{x}{t}\right\rangle & \approx & (1-\sqrt{1-\rho})(a(r+1)-1)+\\
\nonumber & & +\frac{\sqrt{a(1-a)}(1-\sqrt{1-\rho})(1-\rho)(r+1)\cos\varphi}{\sqrt{\rho(1-\rho)}}\\
& & +\frac{r-1}{2}\sqrt{1-\rho}+O(t^{-1}).
\end{eqnarray}

\end{appendix}

%%%%%%%%%%%%%%%%%%%%%%%%%%%%%%%%%%%%%%%%%%%%%%%%%%%%%%%%%%%%%%%%%%%%%%%%%%%%%%%%%%%%%%%%%%%%%%%%%%%%%%%%%%%%%%%%%

\end{document}